6/18/2004

# Complex Systems Analysis of Cell Cycling Models in Carcinogenesis

V. I. Prisecaru and I.C. Baianu

University of Illinois,
Urbana, IL  61801, USA

## Abstract

**A new approach to the modular, complex systems analysis of nonlinear dynamics in cell cycling network transformations involved in carcinogenesis is proposed. Carcinogenesis is a complex process that involves dynamically inter-connected biomolecules in the intercellular, membrane, cytosolic, nuclear and nucleolar compartments that form numerous inter-related pathways referred to as networks. One such family of pathways contains the cell cyclins.  Cyclins are proteins that link several critical pro-apoptotic and other cell cycling/division components, including the tumor suppressor gene TP53 and its product, the Thomsen-Friedenreich antigen (T antigen), Rb, mdm2, c-Myc, p21, p27, Bax, Bad and Bcl-2, which all play major roles in carcinogenesis of many cancers.  This novel theoretical analysis based on recently published studies of cyclin signaling, with special emphasis placed on the roles of cyclins D1 and E, suggests novel clinical trials and rational therapies of cancer through re-establishment of cell cycling inhibition in metastatic cancer cells.**

## Cyclins

Cyclins are proteins that are often overexpressed in cancerous cells (Dobashi et al., 2004). Cyclin-dependent kinases (CDK), their respective cyclins, and inhibitors of CDKs (CKIs) were identified as instrumental components of the cell cycle-regulating machinery. In mammalian cells the complexes of cyclins D1, D2, D3, A and E with CDKs are considered motors that drive cells to enter and pass through the "S" phase. Cell cycle regulation is a critical mechanism governing cell division and proliferation, and is finely regulated by the interaction of cyclins with CDKs and CKIs, among other molecules (Morgan et al., 1995).

The Genetic database I (DSI, Paris, France) Sequence for Cyclin-D1 is found at the PDB website: http://www.dsi.univ-paris5.fr/genatlas/fiche.php?symbol=CCND1

A positive correlation has been noticed between overexpression of several cell-cycle proteins and unfavorable prognoses and outcomes in several different cancer types (van Diest et al., 1995; Handa et al., 1999; Fukuse et al., 2000). In human lung tumors and soft tissue sarcomas, it has recently been discovered that cyclin A/cdk2 complex expression and kinase activity were reliable predictors of proliferation and unfavorable prognosis, thereby further substantiating the epidemiological factors of cyclin signaling (Dobashi et al., 2003; Noguchi et al., 2000). The Cdk2/Cyclin A forms a Complex with an 11-residue recruitment peptide from the Retinoblastoma-Associated protein whose complex structure has been determined by X-ray Diffraction.

## **p21 and p27**

The proteins p21 and p27 are also implicated in cyclin regulation and cancer development (Fig.1). Mouse embryonic fibroblasts that were deficient for p21 and p27 were found to contain less cyclin D1 and D2 (Cheng et al., 1999) as well as cyclin D3 (Bagui et al., 2000) than controls. Similarly, mammary glands of p27-deficient mice were shown to possess decreased cyclin D1 levels (Muraoka et al., 2001). It has been demonstrated *in vivo* that p27 is necessary for maintaining proper levels of cyclins D2 and D3, and this dependency on p27 is common to a wide variety of cells/tissues *in vivo*. The reduction of cyclin D levels under p27-deficient conditions equals the portion of cyclin D pool normally associated with p27 (Bryja et al., 2004). The absence of p27 causes Leydig cells to produce double the testosterone, compared to controls (Bryja et al., 2004), and p27 in Leydig cells appears to serve some other purpose besides regulating the G1/S transition (Bryja et al., 2004).  Regarding the molecular interaction between p27 and D-cyclin, CDK4 is a clear candidate as a mediating molecule (Bryja et al., 2004).

**The Dual Role of D-type Cyclins and p27 in Proliferation and Differentiation**

Cells employ CDK4/6– cyclin D complexes to flexibly titrate p27 from the complexes containing CDK2, and thereby they control their proliferation.  However, mutual dependency between cyclin D and p27 serves also some yet unidentified function in differentiation-related processes.  Thus, loss of p27 not only causes unrestricted growth due to inefficient inhibition of CDK2–cyclin E/A, but may also elicit a decrease in levels of D-type cyclins, resulting in differentiation defects.  Upon ablation of cyclin D, cells lose their ability to titrate p27 from CDK2–cyclin A/E complexes and proliferation is suppressed.  However, defects in differentiation caused by the absence of D-cyclin are reminiscent to defects produced by the absence of p27 (Bryja et al., 2004).  When the changes in levels of p27 and/or D-type cyclins occur, an equilibrium alteration could result between proliferation/differentiation processes that may in the end result in tumorigenesis (Bryja et al., 2004).

# D1 vs. E Cyclins

The D-type and E-type cyclins control the G1 → S phase transition during normal cell cycling and are important components of steroid- and growth factor-induced mitogenesis in breast epithelial cells (Sutherland and Musgrove, 2004).  Cyclin D1 null mice are resistant to breast cancer that is induced by the *neu* and *ras* oncogenes, which suggests a pivotal role for cyclin D1 in the development of some mammary carcinomas (Sutherland and Musgrove, 2004).  Cyclin D1 and E1 are usually overexpressed in breast cancer, with some association with adverse outcomes, which is likely due in part to their ability to confer resistance to endocrine therapies.  The consequences of cyclin E overexpression in breast cancer are related to cyclin E's role in cell cycle progression, and that of cyclin D1 may also be a consequence of a role in transcriptional regulation (Sutherland and Musgrove, 2004).  One critical pathway determining cell cycle transition rates of G1 → S phase is the cyclin/cyclin-dependent kinase (Cdk)/ p16Ink4A/ retinoblastoma protein

(pRb) pathway (Sutherland and Musgrove, 2004). Alterations of different components of this particular pathway are very ubiquitous in human cancer (Malumbres and Barbacid, 2001).

There appears to be a certain degree of tissue specificity in the genetic abnormalities within the Rb pathway. A model relating Rb to cyclin control in the overall scheme of pro-apoptotic behavior is shown below in Fig.1. In breast cancer these abnormalities include the overexpression of cyclins D1, D3 and E1, the decreased expression of the p27Kip1 CKI and p16Ink4A gene silencing through promoter methylation. These aberrations occur with high frequency in breast cancer, as each abnormality occurs in ~40% of primary tumors. This fact implicates a major role for the loss of function of the Rb pathway in breast cancer. Cyclin D1 is the product of the *CCND1* gene and was first connected to breast cancer after localization of the gene to chromosome 11q13, a region commonly amplified in several human carcinomas, including ~15% of breast cancers (Ormandy et al., 2003). The fact that cyclin D1 was overexpressed at the mRNA and protein levels in 50% of primary breast cancers have caused cyclin D1 to be considered one of the most commonly overexpressed breast cancer oncogenes (Gillett et al., 1994; Alle et al., 1998).

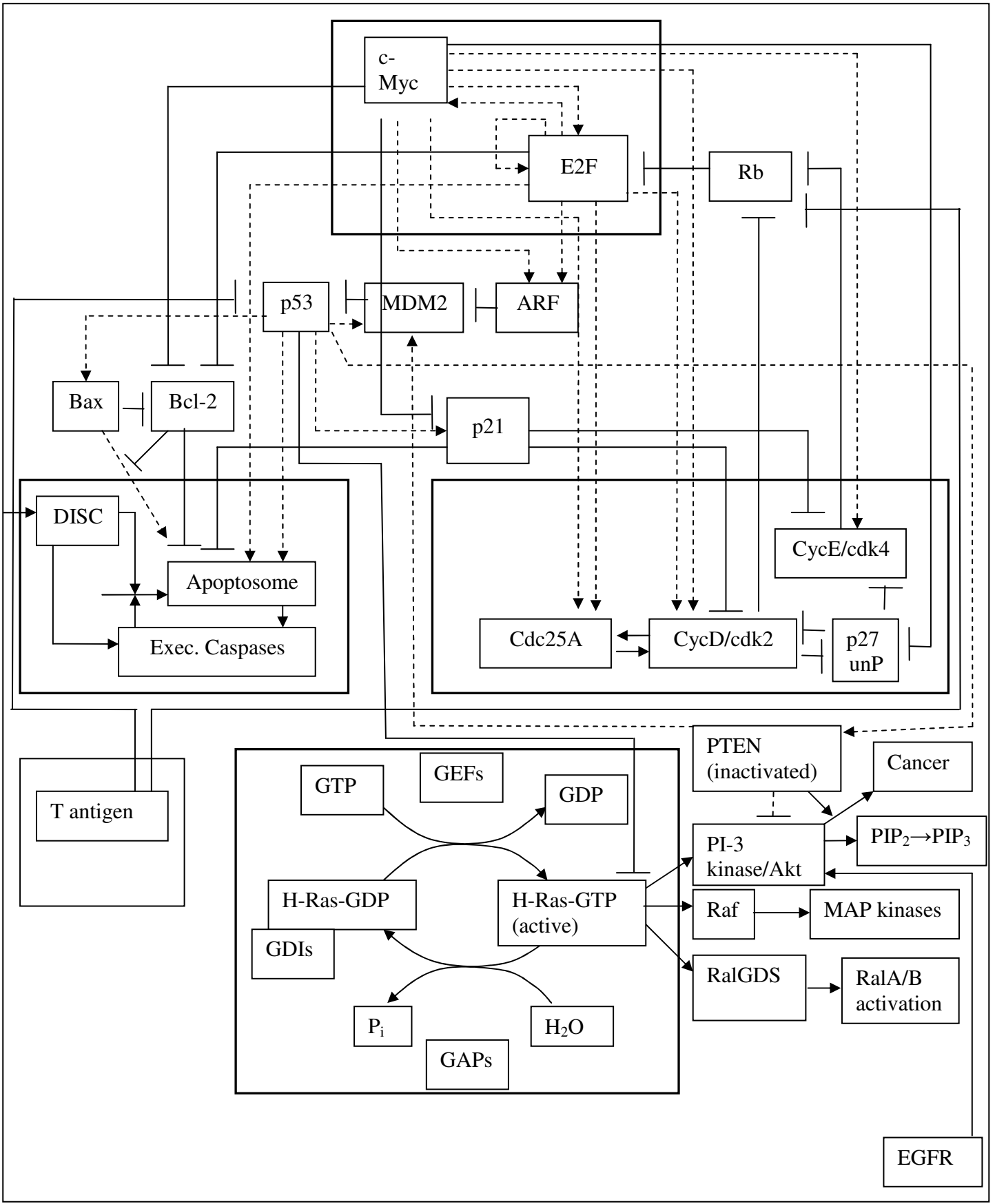

**Fig. 1. Cancer Cell Cycling Model** (substantially modified from Aguda, 2003).

</hr>

Although cyclin E1 locus amplification is rare in breast cancer, the protein product is overexpressed in over 40% of breast carcinomas (Loden et al., 2002). Cyclin D1 is predominantly overexpressed in ERC tumors, and cyclin E overexpression is confined to ER¡ tumors (Gillett et al., 1994; Alle et al., 1998; Loden et al., 2002).

The overexpression of several cell cycle regulators has been strongly associated with apoptotic-like behavior, as well as frank apoptosis, in cancer cells, which include c-Myc, E2F-1 and HPV. Apoptosis and its connection to cell cycle-related proteins is of interest therapeutically, as these types therapies could ultimately lead to the cancer cell annihilation via apoptosis. Recently, a shift has occurred, changing the focus of chemotherapy from exploration of agents that cause cell growth arrest to those that favor apoptosis.

## FGFR Tyrosine Kinases

Fibroblast growth factor receptor (FGFR) tyrosine kinases have recently been studied as they relate to intracellular signaling and their effects on pRb, and are of interest to the field of cancer biology. Overexpression of FGFR tyrosine kinases has been found in many human breast carcinomas and has been associated with poor clinical prognosis (Koziczak et al., 2004). Fibroblast growth factor receptors (FGFRs) are glycoproteins composed of extracellular immunoglobulin (Ig)- like domains, a hydrophobic transmembrane region and a cytoplasmic moiety that contains a tyrosine kinase domain (Koziczak et al., 2004). When active, FGFRs stimulate tyrosine phosphorylation, as well as activation of several signaling molecules: Shc, PI3K, Src, PLCg, Crk, SH2 domain containing phosphatase-2 (SHP-2), p38, STAT1/3 and FGFR substrate 2 (FRS2) (Klint and Claesson-Welsh, 1999). Treatment of tumor cells with the FGFR tyrosine kinase inhibitor leads to a reduction in pRb phosphorylation on serine 795, a site known to be phosphorylated by the cyclin D/cdk4 complex (Koziczak et al., 2004). FGFR signaling may in fact promote cell proliferation by upregulating cyclin D levels. This idea is supported by the fact that ectopic cyclin D1 expression is able to rescue the FGFR inhibitor-mediated antiproliferative effect (Koziczak et al., 2004). Using a cyclin D1 reporter gene, Koziczak et al. (2002) found that

FGFR inhibitor caused a significant reduction in promoter activity, and was reflected in an overall decrease in cyclin D1 mRNA levels.

A recent study employed p27-deficient mice to investigate the significance of p27 for the metabolism of D-type cyclins in differentiated cells (Bryja et al., 2004). The absence of p27 resulted in decreased cyclins D2 and/or D3 levels in several organs. The drop in cyclin D levels that was due to the absence of p27 equaled the amount of cyclin D physically associated with p27 animal controls.

**The Regulation Pathways of p27 Phosphorylation during Cell Cycling can be viewed at the NCI supported CGAP website**:

http://cgap.nci.nih.gov/Pathways/BioCarta/h_p27Pathway).

This indicates the possibility that it is the proportion of p27-associated cyclin D that determines the response to p27 deficiency. Cells in which the D-type cyclin level is dependent on p27 do not up-regulate their CDK2 and CDK4 activities upon deactivation of p27. Moreover, these cells have a negligible amount of p27 bound to CDK2 and/or cyclin A/E under non-cancerous conditions (Bryja et al., 2004). These findings point to the existence of two roles for p27: regulation of the cell cycle through inhibition of CDKs, and participation in the establishment or maintenance of the differentiated status that is achieved in conjunction with D-cyclins (Bryja et al., 2004).

# Ubiquitin

The regulation of protein stability via the ubiquitin–proteasome pathway is critical to the comprehension of the biomolecular basis of cancer development. However, ubiquitin modification of substrates signals many cellular processes (besides proteolysis) that are also important for cancer development. Interestingly, many breast cancer proteins studied by clinical researchers are involved in these specific ubiquitin pathways. These proteins include cyclins, CDK inhibitors and the SCF in cell cycle control, the breast and ovarian cancer suppressor BRCA1-BARD1,

ErbB2/HER2/Neu and its ubiquitin ligase c-Cbl, as well as and the estrogen receptor and its target, Efp.

One function of the ubiquitin–proteasome proteolysis pathway is to label proteins for rapid degradation. It consists of four enzymes: a ubiquitin-activating enzyme (E1), a ubiquitin-conjugating enzyme (E2), a ubiquitin ligase (E3) and the 26S proteasome (Hershko and Ciechanover, 1998). E1 binds to and activates ubiquitin in an ATP-dependent manner through a thiolester bond and then transfers ubiquitin to an E2 enzyme. E2 then transfers ubiquitin to a lysine residue in the substrate via a terminal isopeptide bond through E3. E3 is a scaffold protein that bridges in the substrate and the ubiquitin-bound E2. The resultant covalent bonds of the ubiquitin ligations form polyubiquitinated conjugates that are quickly found and digested by the 26S proteasome. Understanding these pathways may provide many critical clues toward the development of novel diagnostic tools and treatments for cancer patients (Ohta and Fukuda, 2004).

## BRCA1-BARD1 Ubiquitin Ligase

The breast and ovarian cancer susceptibility gene BRCA1 is probably the most studied gene in the breast cancer field because of its clinical significance and multiple functions. The BRCA1 gene encodes a 1863-amino-acid protein (Miki et al., 1994) that consists of a RING-finger domain in its terminal N-region, a region that includes a nuclear localization signal and a domain that binds to many cellular proteins, and tandem BRCT domains in its C-terminal region. BRCA1 is associated with a diverse range of biological processes, such as DNA repair, cell cycle control, transcriptional regulation, apoptosis and centrosome duplication. The only known biochemical function of BRCA1 is E3 ubiquitin ligase activity (Table 1). The N-terminal RING finger domain of BRCA1 interacts with another conformationally similar RING finger protein, BARD1 (Wu et al., 1996; Brzovic et al., 2001), that also contains an N-terminal RING domain and C-terminal BRCT domains (Wu et al., 1996). BRCA1 attains high ubiquitin ligase activity when bound to BARD1 as a heterodimer (Hashizume et al., 2001). Importantly missense mutations in the RING-finger domain of BRCA1 found in familial breast cancer all eradicate the ubiquitin ligase activity of BRCA1-BARD1 (Hashizume et al., 2001; Ruffner et al., 2001; Brzovic et al., 2003). This fact suggests a strong link between BRCA1 ligase activity and its function as a tumor suppressor. The analysis of ubiquitin ligase activity of RING-domain mutations is important not only for the investigation of the biological function of BRCA1, but also to be able to predict a specific patient's propensity for cancer, which may influence the need for prophylactic surgeries.

## Ubiquitin E3 ligases and their targets related to breast cancer

Besides enhancing BRCA1's ubiquitin ligase activity, BARD1 is also critical for BRCA1 stability *in vivo* (Hashizume et al., 2001; Joukov et al., 2001; Xia et al., 2003). Loss of BARD1 leads to a phenotype similar to that of the loss of BRCA1, that is, early embryonic lethality/ chromosomal instability (McCarthy et al., 2003). Moreover, germline mutations of BARD1 are found in breast and ovarian cancer patients (Thai et al., 1998). Although ubiquitin ligase activity may be significant for the role of the BRCA1 gene as a tumor suppressor, the way the activity contributes to BRCA1's biological function remains unknown. Two issues exist that are critical to the elucidation of the role of the BRCA1-BARD1 ubiquitin ligase: the type of polyubiquitin chain built by BRCA1-BARD1 (and its consequences), and the specific identity of its substrates.

In the past decade researchers have identified important functional roles for the D- and E-type cyclins in the evolution of human breast cancers. These genes are among the most commonly overexpressed genes in breast cancer, being overexpressed in the early phases of disease and having proven oncogenic effects on mammary epithelial cells both *in vitro* as well as *in vivo*. Their established role in CDK activation and Rb pathway regulation has directed scientific attention toward aberrant cell cycling as the basis of oncogenic potential. More recent data on the role of different G1 cyclins in the areas of differentiation, chromosome stability and transcriptional regulation indicate that their role in breast cancer is much more complex than initially predicted. Further investigations may yield a more complete understanding of the role of these cyclins regarding the biomolecular basis and pathophysiology of breast cancer, with significant potential benefits clinically, through the identification of novel markers of prognosis and therapeutic responsiveness and potential new targets for innovative clinical intervention.

# Signal Transduction Modulators as Novel Anticancer Drugs

Recently, there has been an increasing number of reports that cancer frequently involves pathogenic mechanisms which give rise to numerous alterations in signal transduction pathways. Therefore, novel therapeutic agents that target specific signal transduction molecules or signaling pathways altered in cancer are currently undergoing clinical trials often with remarkable results in cancer treatments of patients in which chemo- and/or radio- therapy resistant tumors have become established.  For example, a new class of anticancer drugs, the (protein) tyrosine/ threonine kinase inhibitors, such as: STI-571 (**'Gleevec', or Imatinib Mesylate**), ZD-1839 ('Iressa'), OSI-774, and flavopiridol, are ATP-site antagonists and have recently completed phase I and phase II trials. Also directed against signal transduction targets are the monoclonal antibodies Herceptin and C225. Several other kinase antagonists are currently undergoing clinical evaluations, including UCN-01 and PD184352.  Other strategies for downmodulating kinase-driven signaling include 17-allyl-amino-17 demethoxygeldanamycin and rapamycin derivatives. Phospholipase-directed signaling may also be modulated by alkylphospholipids.  Farnesyltransferase inhibitors, originally developed as inhibitors of *ras*-driven signals, may attain activity by affecting other or additional targets. Signal transduction is an efficient method for fine-tuning the development and modeling of cancer treatments. The following detailed background on clinical trial and signal transduction modulators as novel anticancer drugs summarizes the contents of a recent NCI Report (Sausville, Elsayed, Monga and Kim, 2003) and references cited therein.

## Tyrosine Kinase Inhibitors

**STI-571** ('Gleevec' or Imatinib Mesylate) inhibits three kinases:  Abl (all forms), PDGFR and c-kit tyrosine kinases.  It blocks the Bcr-Abl tyrosine kinase, and is important in chronic myelogenous leukemia (CML) patients because CML cells have constitutively active Bcr-Abl tyrosine kinase.  STI-571 *differentially* inhibited the growth of **p210$^{Bcr-Abl}$** CML and **p185$^{Bcr-Abl}$** CML containing acute lymphoblastic leukemia cells and does <u>not</u> affect the normal marrow cells. The effect of STI-571 is exciting because it inhibits <u>c-kit</u>/<u>CD117</u> positive tumors owing to the paucity of interventions for these <u>chemoresistant tumors</u>. For example, significant response in rapidly progressive gastrointestinal tumors (GIST) and soft-tissue sarcomas that were previously resistant to cytostatic, anticancer drugs when Gleevec is not administered simultaneously with such cytostatics.  FDA has approved Gleevec for GIST as well as CML treatments, and is undergoing clinical trials for novel therapeutic strategies of other types of cancer.

**SU5416 is** a ATP-site antagonist of the vascular endothelial growth factor (VEGF) (Flk1/KDR) receptor was designed following studies of the indolin-2-one pharmacophore and the fibroblast growth factor (FGF) receptor tyrosine kinase domain. A Lineweaver-Burk analysis showed SU5416 to be a competitive inhibitor with ATP for the Flk1/KDR and PDGF receptors ($K_i$ 0.16 µM and 0.32 µM, respectively) (Mohammadi et al., 1997; Mendel et al., 2000).  The first SU5416 clinical trial enrolled 63 patients and administered the drug i.v. biweekly (Rosen et al., 1999);  at the higher doses, nausea, vomiting, headache and some liver toxicity were noticed;[ stable disease of greater than 6 months duration was the only reportable outcome in patients with a variety of advanced diseases (colorectal, lung, renal and Kaposi's sarcoma.  Patients with significant progression suffered noticeable increases in vascularity; the occurrence of vascular complications like thrombotic events raises the risk of broad application of this drug (Kuenen et al., 2002).

## Tyrosine Kinase/EGFR Inhibitors

**ZD 1839** ('Iressa').  Epidermal Growth Factor Receptor (EGFR) *activates several downstream signaling pathways* and is overexpressed in numerous types of human cancers, including: non-small cell lung (NSCLC), colorectal, head and neck, bladder, brain, pancreas, breast, ovary, prostate, and gastric cancers (Salomon et al., 1995; Gullick et al., 1991). Overexpression of EGFR is associated with increased invasiveness, resistance to treatment and poor outcomes in several tumor types (Neal et al., 1985; Ke et al., 1998).  Iressa is found to be effective in the treatment of: *Non-small cell lung (NSCLC), colorectal, head and neck, bladder, brain, pancreas, breast, ovarian, prostate and gastric* **cancer types that were previously unresponsive to other chemotherapy** (Salomon et al., 1995; Gullick et al., 1991).  ZD 1839 (Iressa) blocks EGFR; ZD1839 inhibits autophosphorylation, and results in complete regression in some xenograft tumors (Ciardiello et al., 2000; Sirotnak et al., 2000) when used with cytotoxic drugs such as doxorubicin, or in combination with radiation.  Iressa inhibits the **Ras/MAP kinase and**

**STAT-3 transcription factors,** in many tumors**;** the inhibition of the epidermal growth factor receptor (EGFR) has been of significant interest lately, partially because of the autocrine activation of EGFR and several downstream pathways, such as the ras/MAP kinase and STAT-3 transcription factors, in several tumors. The activated EGFR pathway *induces entry into the cell cycle, inhibition of apoptosis, and also activation of angiogenesis and motility*. Several phase I and II studies with Iressa have already been completed (Ferry et al., 2000; Negoro et al., 2001; Baselga et al., 2000). Daily oral doses have ranged from 50 to 700 mg for 2 to 4 weeks. ZD1839 resulted in some responses in NSCLC and prostate cancer, and stability of disease (over 4 months) in several patients (Ferry et al., 2000; Negoro et al., 2001; Baselga et al., 2000). 22% of Japanese patients achieved partial response (Negoro et al., 2001). Side effects have been relatively mild and have included diarrhea and rash.

**OSI-774**, Erlotinib, or 'Tarceva' is also an EGFR inhibitor; it binds very tightly to EGFR, causing EGFR inhibition. It produces downstream inhibition of the P13/MAPK signal transduction pathways, **resulting in accumulation of p27, that leads to cell cycle arrest at the G1 phase and induction of apoptosis** (Moyer et al., 1997). EGFR-TK is more than 1000 fold sensitive to 'Tarceva' compared with any other tyrosine kinases. Therefore, it is a very specific inhibitor of EGFR –TK and **reduces very markedly the phosphorylated EGFR-TK.** The $IC_{50}$ for Tarceva is 2 nM (when measured by purified EGFR-TK inhibition in biochemical assays), and its value is 20 nM for the EGFR-TK autophosphorylation when measured in intact cells. The proposed mechanism of action: reversible inhibition of EGFR-TK through competitive binding to the ATP site. Results of preliminary Clinical Trials include: partial responses in patients with colorectal cancer and renal cell carcinoma (kidney), as well as > 5 month stabilization in: colon, prostate, cervical, NSCLC and head and neck cancers.

**Herceptin** (Trastuzumab), a recombinant humanized monoclonal antibody directed against HER2 is known as 'Herceptin' (Carter et al., 1992). The HER2/neu gene increases the kinase activity, initiating signal transduction, leading to proliferation and differentiation in approximately 30% of human breast cancers (up to 50 to 100 gene copies/cell). The HER2/neu gene makes a type I receptor tyrosine kinase encoding a 185 kDa surface membrane receptor protein. Phase I trials showed that the dose of trastuzumab (i.v. 10 to 500 mg single dose or weekly) could be increased without toxicity and that pharmacokinetics were dose-dependent (Shak et al., 1999). Phase II trials response is > 5.3 months. Phase III trial patients received doxorubicin or epirubicin plus cyclophosphamide, and 28% of patients treated with chemotherapy and trastuzumab were free of tumor progression, compared with 9% of the patients treated with chemotherapy alone. The monoclonal antibody of the membrane receptor HER2 signaling protein is much more efficient than chemotherapy alone. About 1 in 5 of the patients had cardiac dysfunction, where trastuzumab was at 4 mg/kg body weight initially. A phase II trial was conducted with 46 HER2 (+) metastatic breast cancer patients who had failed prior cytotoxic chemotherapy (Baselga et al., 1996). Objective responses were seen in 5 of 43 assessable patients, including 1 complete remission and 4 partial remissions. A second phase II trial (Pegram et al., 1998) combined trastuzumab with cisplatin in 39 HER2 (+) metastatic patients who had failed prior chemotherapy. Of the 37 subjects, 9 achieved a partial response and 9 had a minor response or stability. A randomized, placebo-controlled phase III study was performed to determine efficacy and safety of adding trastuzumab to chemotherapy in breast carcinoma. 28% of patients treated with both were disease progression-free at 12 months, compared with 9% of the patients treated only with chemotherapy. The treatment is indicated as a single agent for patients that have failed earlier therapy, and it is used also as first-line treatment for metastatic disease when used in combination with paclitaxel. *Trustuzumab has been approved* by the FDA for use in women with metastatic breast cancer with HER2-positive tumors**.**

**Cetuximab**. An antibody-based approach to affecting tyrosine kinase signaling is by cetuximab, a humanized monoclonal antibody against the EGFR. MAb225, a murine monoclonal antibody that specifically binds to EGFR, specifically competes with signal transduction initiated by TGF-$\alpha$ (Gill et al., 1984). Cetuximab (C225) is the human-mouse chimeric version of Mab225, which specifically binds to the EGFR with high affinity, preventing the ligand from interacting with the receptor. Preclinical studies show that cetuximab results in cell-cycle arrest as well as apoptosis in different contexts (Huang et al., 1999; Peng et al., 1996). A synergistic effect of cetuximab with cytotoxic chemotherapy has been seen with cisplatin, doxorubicin (Baselga et al., 1993), gemcitabine (Bruns et al., 2000), docetaxel (Tortora et al., 1999), and paclitaxel (Inoue et al., 2000). Early phase I trials demonstrated that cetuximab displays nonlinear, dose-dependent pharmacokinetics that are not altered by coadministration of cisplatin (Baselga et al., 2000). These studies were conducted in patients with tumors overexpressing EGFR. There were only 5 episodes of severe C225-related toxicities among the 52 patients. Two patients with head and neck tumors who received cetuximab at doses of 200 mg/m$^2$ and 400 mg/m$^2$ with cisplatin exhibited a partial response. In light of these results, the clinical development of cetuximab is continuing with a number of phase II and III studies.

**Serine-Threonine Kinase Antagonists (STKAs)**

      **Rapamycin Congeners.** Rapamycin (Sirolimus, Rapamune) is a macrolide fungicide that binds intracellularly to the immunophilin FKBP12, and the resultant complex inhibits the activity of a 290-kDa kinase known as mammalian target of rapamycin (mTOR). Rapamycin is isolated from the bacteria *Streptomyces hygroscopicus* and is found to have potent antimicrobial and immunosuppressive properties (Baker et al., 1978). Sirolimus was approved by the FDA for prevention of allograft rejection after organ transplantation (Sehgal et al., 1995). Further studies with rapamycin revealed significant antitumor activity (Eng et al., 1984). This is understandable given the importance of mTOR in mitogenic cell signaling. mTOR is a kinase member of PI3K-related kinase family that is activated in response to growth signaling through the PI3K/Akt pathway. Activation of mTOR results in increased translation of several critical cell-cycle regulatory mRNAs through two downstream effector kinases, p70S6K and 4E-BP1/PHAS (Sekulic et al., 2000; Gingras et al., 1998). Rapamycin causes $G_1$ cell-cycle arrest by increasing the turnover of cyclin D1 (Hashemolhosseini et al., 1998), preventing upregulation of cyclins D3 and E (Decker et al., 2001), upregulating p27$^{KIP1}$, and inhibiting cyclin A-dependent kinase activity (Kawamata et al., 1998). Several analogs of rapamycin have been selected for further development as anticancer agents. CCI-779, an ester of rapamycin, has significant antiproliferative effect and favorable toxicology profile and is being studied in several phase I trials in humans (Hidalgo et al., 2000; Raymond et al., 2000). Several partial responses have been documented in renal cell carcinoma, NSCLC, neuroendocrine tumors, and breast cancer, in addition to minor responses or stable disease in several tumor types (Hidalgo et al., 2000; Raymond et al., 2000). RAD001, an orally bioavailable hydroxyethyl ether derivative of rapamycin, also has potent activity against various animal xenograft models of human tumors; an antiangiogenic effect may account in part for its antiproliferative properties (O'Reilly et al., 2002).

**MEK Inhibitor, PD 184352**.  The stimulation of *Ras-mediated signal pathways* results in a cascade of downstream kinase activation including Raf, which phosphorylates two distinct serine residues on the dual-specificity kinase MEK (MAP kinase-kinase) (Lewis et al., 1998).  MEK, in turn, activates and exclusively phosphorylates two subsequent kinases, ERK1 and ERK2 (MAPK), on specific tyrosine and threonine residues within each kinase. These kinases phosphorylate a variety of substrates including transcription factors critical to cell proliferation and tumor invasion (e.g., Marais et al., 1993).  In cytotoxicity studies, correlation between sensitivity to PD184352 and increased activated MAPK levels was observed in some cells—in particular, colon cancer cells. Higher levels of MAPK activation were observed in colon tumor tissue versus normal mucosa as this event occurs late in colon carcinogenesis (Sebolt-Leopold et al., 1999).  In mice with colon 26 xenograft model treated with PD184352, excision and assay of tumor cells revealed diminished phospho-MAPK levels. After drug withdrawal, a return to baseline levels was observed reflecting the cytostatic nature of the inhibition. The pharmacodynamic measurement of activated MAPK in tumor tissue may be used as a biological marker of drug activity as antibodies specific for phosphorylated MAPK are available.

**Bryostatin.**  The bryostatins represent a large family of secondary metabolites produced in extremely small amounts by the marine invertebrate, *Bugula neritina* of the phylum *Ectoprocta* (Pettit et al., 1991). The various bryostatins are distinguished by varying side chains off the macrocyclic lactone ring structure. Despite this close structural relationship, these nontumor-promoting PKC activators have different biologic activities and spectrum of toxicity (Kraft et al., 1996; Jones et al., 1990). Bryostatin 1 (Bryo 1) is the prototype of this 17-member family and the most extensively studied in humans. Initial isolation of Bryo 1 was based on its antineoplastic activity against the murine P388 lymphocytic leukemia.  Bryo 1 is a potent and rapid activator of PKC; however, unlike other PKC activators, including phorbol myristate acetate (PMA), Bryo 1 lacks tumor-promoting capabilities.  The first two published phase I trials evaluated Bryo 1 administered as a 1 h intravenous infusion (Prendiville et al., 1993; Philip et al., 1993).  The DLT was myalgia, occurring approximately 48 h after treatment and lasting up to several weeks at the highest dose levels (65 µg/m$^2$/dose).  The MTD was 50 µg/m$^2$, and the recommended dose for phase II trials was 35 to 50 µg/m$^2$ every two weeks. Partial responses were observed in two patients with malignant melanoma, which lasted 6 months and 10 months.  Plasma levels of tumor necrosis factor-alpha (TNF-α) and interleukin-6 (IL-6) increased 2 h and 24 h after treatment, respectively, and were dose related.

**UCN-01** (7-OH Staurosporine): Staurosporine, a natural product isolated from *Streptomyces staurosporeus*, is a relatively broad, nonspecific protein kinase antagonist, originally isolated in an effort to define inhibitors of protein kinase C (PKC). 7-OH staurosporine (UCN-01) was defined as a more selective, but not specific, PKC antagonist. Two prominent effects of UCN-01 have emerged in preclinical studies in vitro: induction of cell-cycle arrest, and abrogation of the checkpoint to cell-cycle progression induced by DNA damaging agents. UCN-01 inhibited cell growth in several in vitro and in vivo human tumor preclinical models (Akinaga et al., 2000); however, antiproliferative activity on the part of UCN-01 cannot be explained solely by inhibition of PKC. First, in cell-cycle analyses UCN-01 inhibits $Rb^+$ cells at G1/S phase of the cell cycle (Akiyama et al., 1997). In addition, cells treated with various concentrations of UCN-01 showed *decreased pRb phosphorylation in a dose-dependent manner* (Chen et al., 1999). These results suggest that CDK2- or CDK4-regulated steps are targets for UCN-01-induced cell-cycle arrest. UCN-01 abrogates the DNA damage-induced checkpoints to cell-cycle progression in G2 (Bunch et al., 1996; Wang et al., 1996) and in S phase (Shao et al., 1997). It is noteworthy that these effects were apparent at drug concentrations that appeared to have little direct effect on cell proliferation or that caused enhanced cytotoxicity by clonogenic or proliferation assays. In addition, they provided a mechanistic framework for prior observations that DNA-damaging agents such as mitomycin (Akinaga et al., 1993) could greatly potentiate UCN-01 action. In contrast to animal studies, UCN-01 displayed strong binding to human plasma proteins, apparently to the α1-acid glycoprotein (AAG) in initial human phase I clinical trials (Sausville et al., 2001; Fuse et al., 1998). One partial response occurred in a patient with melanoma, and a protracted (>4 year) period of stabilization of minimal residual disease was observed in a patient with *alk*(+) anaplastic large cell lymphoma.

Miltefosine and Perifosine (ALP Analogs). Certain alkylphospholipids (ALP) (e.g., Rac-1-O-octadecyl-2-O-methyl-glycero-3-phosphocholine [ET-18-$OCH_3$, edelfosine]) when given to mice prior to transplantation of Ehrlich ascites carcinoma cells, effectively prevent growth of this tumor (Tarnowski et al., 1978). Enhancement of immune defense against tumor cells was initially considered a plausible mechanism and has been demonstrated on multiple occasions by a number of ALP analogs. Edelfosine is also able to induce apoptosis in HL60 leukemic cells, even in low concentrations and after short incubation times. In U937 leukemic cells, the compound induced apoptosis rapidly, whereas in epithelial HeLa tumor cells this induction required prolonged times of treatment (Mollinedo et al., 1993). All ALP analogs studied so far cause an indirect inhibition of PKC, most likely as a result of the reduced formation of diacylglycerol through inhibition of phospholipase C (Uberall et al., 1991; Seewald et al., 1990). Additional antiproliferative mechanisms could involve altered growth factor receptor function, as well as recent evidence of *p21 induction by an as yet undefined pathway* (Patel et al., 2002) *irrespective of p53 function.* Eight phase I-II studies, consisting of 443 patients using topically applied miltefosine 2%-8% for skin metastases in patients with breast cancer, showed a median response rate of 38% (Unger et al., 1988; Smorenburg et al., 2000; Terwogt et al., 1999). Evidence from the trials led to the approval of miltefosine, licensed as Miltex©, in Germany for the treatment of cutaneous breast cancer and cutaneous lymphomas. The heterocyclic alkylphosphocholine derivative octadecyl-(1,1-dimethyl-piperidino-4-yl) phosphate (D-21266; perifosine) was developed and selected for improved gastrointestinal tolerability. A number of phase I studies are presently ongoing in Europe and the United States; early evidence points to better tolerability and less gastrointestinal toxicity (Messmann et al., 2001).

**Proteosome Inhibitor PS-341.** The proteasome, a multicatalytic protease responsible for degradation of most proteins with the cell, has emerged as a new target for anticancer drug development. The 20S proteasome is involved in the degradation of several cell-cycle regulatory proteins such as cyclins (A, B, D, E), cyclin-dependent kinase inhibitors ($p21^{WAF1/CIP1}$ and p27), oncogenes (c-fos/c-jun, c-myc, N-myc), and p53 and regulatory proteins (I$\kappa$B, p130) (King et al., 1996). Inhibition of the 20S proteasome pathway aims at altering the cell cycle to promote apoptosis (An et al., 2000). Although the proteasome is present in all cells, transformed and dividing cells are most sensitive to its inhibition (Drexler et al., 1997). PS-341 is the first proteasome inhibitor to enter human trials. It is a boronic acid dipeptide that specifically inhibits the 20S proteasome presumably through the stability of a boron-threonine bond that forms at the active site of the proteasome. It was found to have substantial cytotoxicity against a wide range of human tumor cells in the NCI 60 cell line anticancer drug screen (Adams te al., 1999). PS-341 causes accumulation of cyclin A, cyclin B, $p21^{WAF1/CIP1}$, and wild-type p53 and arrests the cells at the S and G$_2$/M phases followed by nuclear fragmentation and apoptosis. PS-341 significantly inhibited NF-κB DNA binding and functional reporter activity (Sunwoo et al., 2001). Several phase I studies evaluated various schedules of PS-341 administration. At the MTD recommended for phase II studies (1.25 mg/m$^2$-1.3 mg/m$^2$), a 65%-72% inhibition of 20S proteasome was achieved (Erlichman et al., 2001; Aghajanian et al., 2001). An average 54% inhibition of proteasome was achieved in patients' tumors (Hamilton et al., 2001). In these phase I studies several patients achieved partial responses and disease stabilization including a bronchoalveolar NSCLC, melanoma, sarcoma, lung adenocarcinoma, and malignant fibrous histiocytoma. Patients usually had more toxicity with the second cycle of treatment. Currently several phase II clinical trials are evaluating PS-341 as a single agent in hematologic malignancies, neuroendocrine, renal cell, melanoma, breast, brain, pediatric tumors, and several other solid tumors. Significant antitumor effects were documented in a phase II study of PS-341 in refractory multiple myeloma (Richardson et al., 2001).

## Farnesyl Transferase Inhibitors

Ras genes are mutated in 30% of all human cancers with K-Ras being the most common. This family of genes encodes GTP binding proteins important in malignant transformation, cell growth, and intracellular signal transduction. Normal ras binds GTP and in the GTP-bound state interacts with numerous effectors including the *raf* proto-oncogene kinase and phosphatidyl-inositol 3-kinase. Three isoforms, Harvey(Ha), Kirsten(K), and N-isoforms have been described, with mutation of the GTPase of the K isoform resulting in a persisting signaling capacity in approximately 20% of human epithelial tumors. N-ras is mutated in a smaller proportion of malignancies, predominantly leukemias. Ras function requires lipophilic anchorage to the cell membrane by lipid prenylation. This requires posttranslational modification or covalent thioether bond formation between a farnesyl group (C15) and a cysteine residue at the ras carboxy terminus. A "GTT shunt pathway" maintains K-Ras in an active prenylated, membrane-bound form and explains in part the requirements for higher farnesyl transferase inhibitor (FTI) dose or cotreatment with a GTT inhibitor for significant growth inhibition in K-Ras models (Pendergast et al., 2000). Several classes of FTIs have been developed in an initial effort to define inhibitors of Ras function and, in general, compete with the enzyme substrates, the CAAX tetrapeptide, and farnesyl pyrophosphate (FFP). The CAAX competitors are generally peptidomimetic agents that mimic the carboxy terminal portion of the Ras protein.

**SCH66336** is a novel oral agent derived from a class of nonpeptide, nonthiol-containing, CAAX mimetic FTIs (Bishop et al., 1995). The drug inhibits in vitro FT activity with an $IC_{50}$ of 1.9 nM for H-ras, 2.8 nM for N-ras, and 5.2 nM for K-ras. Inhibition of cells with activated ras and anchorage-independent growth was noted with $IC_{50}$ 75 nM in H-ras versus 400 nM with K-ras-driven cells (Liu et al., 1999). The observed growth inhibition of tumor cells in soft agar and in xenografts was independent of ras mutational status because even wild-type ras cells were sensitive (Liu et al., 1998). The phase I experience with SCH66336 involved 20 patients using a twice a day schedule over 7 days every 21 days. Eight patients had stable disease, and treatment for up to 10 cycles was possible in a few patients. Antitumor activity was reported in one patient with advanced NSCLC who had a greater than 50% reduction in an adrenal metastasis and received treatment for 14 months (Adjei et al., 2000).

**R115777** is a substituted quinolone and competitive inhibitor of the CAAX peptide binding site of FT (End et al., 1999). The compound inhibits in vitro K-Ras farnesylation ($IC_{50}$ 7.9 nM) and exerts antiproliferative effects in cell lines such as H-Ras-transformed fibroblasts ($IC_{50}$ 1.7 nM) and K-Ras-driven colon and pancreatic cells lines (at roughly $IC_{50}$ 20 nM) (End et al., 2001). The initial clinical experience with R115777 in 27 patients was reported by Zujewski et al. (Zujewski et al., 2000). Of the 27 patients treated, 8 had stable disease after 3 treatment cycles, and 4 patients continued treatment with the longest reaching 5 months. A patient with metastatic colon cancer had symptomatic improvement and a 50% reduction in carcinoembryonic antigen (CEA) levels. In another phase I study with R115777, two patients with stable disease exceeding 6 months were reported. A partial response in a NSCLC patient lasting 4 months was reported (Schellens et al., 2000). Finally, 3 advanced breast cancer patients treated continuously at 300 mg twice a day attained confirmed partial responses whereas another 9 patients had stable disease of at least 3 months duration (Johnston et al., 2000).

A very interesting outcome was obtained in patients with myelodysplastic syndrome or relapsed or poor prognosis leukemias, where a phase I dose escalation study revealed DLT at 1200 mg twice per day, consisting of neurotoxicity, with non-DLTs including renal insufficiency and myelosuppression. There was clear evidence of downmodulation of erk kinase activity, along with the farnesylation status of lamin A and HDJ-2. Clinical responses occurred in 29% of 34 evaluable patients, including 2 complete responses (Karp et al., 2001). Though there were no mutations in N-Ras detected in this patient population, this study did suggest that in addition to clinical activity there was some evidence of downmodulation of signaling as well as farnesylation-directed activities.

# Conclusions

## Specific Results on the use of Signal Transduction Modulators as Novel Anticancer Drugs

1. Gleevec, Herceptin (Trastuzumab) and CCI-779 have been approved by FDA for treatments of several types of cancer.
2. Preclinical studies with C225 showed that 'Cetuximab' results in cell cycle arrest, as well as apoptosis in several types of tumors, and it had synergistic effects with cytotoxic chemotherapy; Cetuximab (C225) was reported to be tested further in Phase II and III clinical trials.
3. Flavopiridol : The goal would be to develop new STKAs that would be similar to flavopyridol, or HMR 1275. Some STKAs act as blockers of Cyclin D1, therefore causing cycle arrest by direct transcription repression of cyclin D1 mRNA (79), and in mantle cell lymphoma, flavopyridol delayed significantly progression of disease in 84% of patients (80).Cytostasis effects are significant and were observed with the flavopyridol to colorectal and prostate carcinoma xenograft models.

## Evaluation of Results and Related Developments

The results obtained so far in clinical trials convey the promise as well as the challenges encountered in developing signal transduction inhibitors (STIs) for significant improvements of cancer treatment efficacy with such new anticancer drugs. These STI molecules represent a marked departure from prior chemotherapeutic approaches that have been, and still are, mostly based on cytotoxic anticancer drugs. The fact that encouraging responses with a number of novel STIs have been seen in cancer patients-- as expected from mechanistic *in vitro* experiments-- reaffirms the relevance of Integrated Cancer Biology research in charting the future course of cancer developmental therapeutics, and especially the importance of understanding signaling transduction functions and the underlying biodynamics in signaling pathways of cancer cells. On the other hand, the initial results obtained raise a number of issues that should be considered before the field can advance towards the eradication of cancer. One such major issue is the lower response observed clinically in cancer patients with some of the new STIs compared to the expected value. Whereas some anticancer agents have entered initial clinical trials with extensive efforts to document target-based effects in conjunction with pharmacology and clinical toxicity evaluations, other agents have not done so, and in those instances the phase I study has no depth, and does not provide valuable information. Therefore, lacking clear evidence of definite clinical response, one cannot confidently move forward to the next trail phase.

Effective, rational design of combinations with standard cytotoxic agents also remains a challenge in the absence of basic information on anticancer drug interactions. Preclinical models of synergistic effects with signaling agents often proceeds from empiricism rather than understanding on a mechanistic basis, which is what should guide clinical implementation. Another important aspect that must be considered is the need for accurate means of diagnosing the dependence of a tumor on a particular signaling pathway, or target, for novel therapeutic strategies. All of problems and drawbacks considered here call for renewed efforts to define improved assays of target effects in the preclinical phase of a drug's development that can also be translated to the clinical arena. Microarray, proteomic and interactomic, approaches offer the promise of providing such means (Mohr, S. et al., 2002) but *these must also be integrated into the clinical trials process*.

The vast amounts of data becoming available from such high throughput measurements with microarrays on human tumors and tissues also require the employment of powerful computers, effective algorithms for data computational analysis, as well as major advances in our understanding of the **complex systems dynamics of cancer cells**, malignant tumors and tumor-immune interactions in human subjects. Furthermore, higher sensitivity, faster and less expensive diagnostic means must be developed and tested in the clinic to establish their applicability to different types of cancer, and also for the early detection of cancer. Molecular profiling and selective molecular imaging of cancer in human subjects are two such closely related diagnostic means that would make possible *individualized* cancer therapy, as well as provide the ability to proceed with economically feasible screening of human populations that are considered to be at risk for specific types of cancer by utilizing relatively inexpensive and non-invasive means for early detection of cancer. NIR and fluorescence microspectroscopy based techniques are two such very promising means for the early detection of cancer, as well as for the effective monitoring of the cancer therapy effects in patients, thus also providing the accurate means for diagnosing the dependence of a tumor on a particular signaling pathway, or target, for novel therapeutic strategies.